\documentclass[twocolumn,epjc3]{svjour3}  
\RequirePackage{graphicx}
\RequirePackage[utf8]{inputenc}
\usepackage{amsmath}	
\usepackage{amssymb}	
\usepackage[dvipsnames]{xcolor}
\usepackage{tikz}
\usetikzlibrary{patterns}
\usepackage{tkz-euclide}
\usepackage{ulem}

\RequirePackage[T1]{fontenc}

\RequirePackage[numbers,sort&compress]{natbib}
\RequirePackage[colorlinks,citecolor=blue,urlcolor=blue,linkcolor=blue]{hyperref}
\journalname{Eur. Phys. J. C}
\begin{document}\sloppy
\title{A quest for the origin of the Sagnac effect}


\author{Arunava Bhadra\thanksref{addr1,e1}
        \and
        Souvik Ghose\thanksref{addr2,e2} 
				\and
				Biplab Raychaudhuri\thanksref{addr3,e3}
}

\thankstext{e1}{e-mail: aru\_bhadra@yahoo.com}
\thankstext{e2}{e-mail: souvikghose@hri.res.in}
\thankstext{e3}{e-mail: biplabphy@visva-bharati.ac.in}


\institute{ High Energy $\&$ Cosmic Ray Research Center, University of North Bengal, Siliguri 734013, India\label{addr1}
           \and
           Department of Physics, Harish-Chandra Research Institute, Chhatnag Road, Jhunsi, Allahabad (Prayagraj), 211019, Uttar
Pradesh, India. \label{addr2}
           \and
					Department of Physics, Visva-Bharati University, Santiniketan, West Bengal, India, 731235 \label{addr3}
}

\date{Received: date / Accepted: date}

\maketitle

\begin{abstract}
In the literature, there is no consensus on the origin of the relativistic Sagnac effect, particularly from the standpoint of the rotating observer. The experiments of Wang et al. \cite{wang2003modified,wang2004generalized} has, however, questioned the pivotal role of rotation of the platform in Sagnac effect. Recently, the relative motion between the reflectors which force light to propagate along a closed path and the observer has been ascribed as the cause of the Sagnac effect. Here, we propose a thought experiment on linear Sagnac effect and explore another one proposed earlier to demonstrate that the origin of the Sagnac effect is neither the rotation of frame affecting clock synchronization nor the relative motion between the source and the observer; Sagnac effect originates purely due to asymmetric position of the observer with respect to the light paths. Such a conclusion is validated by analysis of a gedanken Sagnac kind experiment involving rotation. 

\keywords{Sagnac effect \and STR \and linear  }

\end{abstract}


\section{Introduction}Sagnac effect is the difference in phase (or time of arrivals) of two coherent light beams (originated from a single light beam) propagating along a rotating closed loop in opposite directions. The effect was first discovered by Sagnac \cite{sagnac1913regarding,sagnac1914effet}. The effect has been observed experimentally in a wide range of wavelength bands, from radio to x-rays as well as using matter waves \cite{vavilov1956eksperimental,post1967sagnac,frankfurt1972optics,hasselbach1993sagnac,pascoli2017sagnac}. The Sagnac effect receives a lot of interest owing to its practical use in Global Positioning System \cite{ashby2004sagnac}, fiber optic gyroscope, ring laser gyroscope, etc. which are essentially Sagnac interferometers \cite{rodloff1985laserkreisel,schreiber2013invited}. The effect also has applications in geodesy and seismology \cite{velikoseltsev2010sagnac}. The Sagnac interferometer is nowadays used as a tool in civil aviation \cite{rodloff1985laserkreisel,schreiber2013invited,velikoseltsev2010sagnac}, to test gravitation theories \cite{scully1981proposed,schiff1960motion,bosi2011measuring,tartaglia2004sagnac,bhadra02}, to examine quantum properties \cite{ashby2016observation}. \\

The Sagnac effect, which is generally considered as one of the basic experimental effects of STR, is a first-order kinematic effect in $v/c$. Sagnac, however, proposed the experiment in support of his ether model and explained the effect without using relativity theory. Since then numerous studies have been conducted on the interpretation of the effect, particularly from the viewpoint of the (co-)rotating observer \cite{malykin1997earlier} which include standard special relativistic description \cite{laue1920versuch}, general relativistic  description \cite{langevin1921theorie,landau1975classical,tonnelat2012principles,ashtekar1975sagnac, benedetto2019general}, synchronization issues in rotating frame \cite{selleri1997relativity,goy1997time,croca1999one}, violation of relativity in rotating frame \cite{sagnac1914effet,kelly1995time} etc. Most of such approaches yield the correct magnitude of the Sagnac effect despite their vast differences in the physical basis of the effect. \\

In the early part of the present century a couple of experimental studies have demonstrated that a Sagnac like effect, which the authors called the generalized Sagnac effect, occurs in a light waveguide loop consisting of both linearly and circularly moving segments \cite{wang2003modified,wang2004generalized}. Interestingly, it was found that both the segments contribute to the generalized Sagnac effect which implies that rotational motion is not essential for the Sagnac effect. Sagnac type gedanken experiment with linear motion (``Linear Sagnac Effect''(LSE)) was probably first proposed by Ghosal et al., \cite{grcs} and they found that the Sagnac delay in the linear case perfectly matches that of the original Sagnac effect when the result is written in terms of the linear velocity without using the area enclosed by the circuit. 

Noting the experimental findings of Wang et. al. \cite{wang2003modified,wang2004generalized} 
recently, Tartaglia and Ruggiero \cite{tartaglia2015sagnac} 
have proposed a Sagnac-like thought experiment consisting of a rectangular closed light path in which an observer is moving along a side of the rectangle with a linear velocity. They concluded that Sagnac effect occurs on the closed path of light in space and when a relative motion exists between the observer and the physical device (mirrors) restricting light to propagate along a closed path.

An important point to be noted at this stage is that in the original Sagnac experiment, both the observer and the physical system employed to move light along the circular path rotate with the same angular velocity. Hence, the existence of a relative motion between the observer and the mirrors~\cite{tartaglia2015sagnac} cannot be the root cause of the Sagnac effect. In this work, we argue that the cause of the Sagnac or generalized Sagnac effect is the non-mid point measurement of arrival times of the counter-propagating light rays (causing unequal path lengths traversed by the light rays in reaching the interferometer), both in the lab frame and the Interferometer frame.

\begin{figure}
\begin{center}
	\begin{tikzpicture}[scale=0.75]
	\fill[pattern=north west lines, pattern color=blue] (0,0) rectangle (.5,2); 
	\draw (.5,0) -- (.5,2);
	\draw[red, thin, dashed] (7.6,0) -- (7.6,2);
	\draw[green, thin, dashed] (8.7,0) -- (8.7,2);
	\begin{scope}[xshift=10cm]
	\fill[pattern=north west lines, pattern color=blue] (0,0) rectangle (.5,2);
	\draw (0,0) -- (0,2);
	\end{scope}
	\begin{scope}[yshift=1.5cm]
	\draw[<-] (7.6,0) -- (9.9,0);
	\end{scope}
	
	\begin{scope}[yshift=1.2cm]
	\draw[->] (.7,0) -- (9.9,0);
	\end{scope}
	\begin{scope}[yshift=0.9cm]
	\draw[<-] (.7,0) --  (5,0);
	\draw[thick, red, ->] (6.2,0)--(8.7,0);	
	\end{scope}
	\begin{scope}[yshift=0.6cm]
	
	\node[] at (7.4,0) {$D_1$};
	\node[] at (8.7,0) {$D_2$};
	\node[red] at (8,0) {$v$};;
	\node[] at (5,0) {$O$};
	\node[] at (6.2,0) {$Q$};
	
	\end{scope}
	\node[below] at (.2,0) {$B$};
	\node[below] at (10.2,0) {$A$};
	\begin{scope}[yshift=1.05cm]	
	\draw[blue, dashed, ->] (5,0) -- (9.9,0);	
	\end{scope}
	\begin{scope}[yshift=1.3cm]	
	\draw[blue, dashed,<<-] (0.6,0) -- (9.9,0);	
	\end{scope}
	\begin{scope}[yshift=1.6cm]	
	\draw[blue, dashed,->] (0.6,0) -- (8.7,0);	
	\end{scope}
	\end{tikzpicture}
\end{center}
\caption{Linear Sagnac}
\label{fig:linear}
\end{figure}
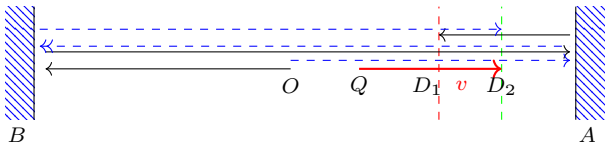

\section{A Thought Experiment}
Let us consider a simple gadanken experiment where two light beams, originating from a single one, using beam splitter, are allowed to propagate in two opposite directions $OA$ and $OB$ (fig. \ref{fig:linear}) along closed linear paths in the lab frame. For convenience, we choose our coordinate system in such a way that the points $A, O, B$ are on the $x$-axis. $O$ is the midpoint of $AB$ so that $OA=OB=L$. We shall take different situations involving the dynamics of the detector and the reflecting mirrors which are employed to restrict the motion of the light rays in closed paths. The coordinates of the frame of reference attached with the detector/interferometer are denoted with primes as superscript. 

In the first case let us consider the detector ($I$), which was initially (i.e. at the time light rays start propagating along $OA$ and $OB$) at point $Q$ where $OQ=q$, moves with a linear constant velocity $v$ with respect to the lab frame along positive $x$ direction. The mirrors (reflectors), which are initially at $A$ and $B$, also move with the same velocity $v$ along the same direction so that no relative velocity exists between the observer attached to the interferometer and the mirrors, as in the original Sagnac effect. In the detector frame, the difference in arrival of time between the lights traveling the path $O^{\prime}A^{\prime}O^{\prime}B^{\prime}Q^{\prime}$ and $O^{\prime}B^{\prime}O^{\prime}A^{\prime}Q^{\prime}$ will be 

\begin{equation}
\Delta t= \frac{2q}{\gamma c}, 
\end{equation}

where $\gamma \equiv 1/\sqrt{1-v^2/c^2}$ is the Lorentz factor. When q=0, i.e. when the detector at the mid point of the light path $O^{\prime}A^{\prime}O^{\prime}B^{\prime}Q^{\prime}$ $(O^{\prime}A^{\prime}=O^{\prime}B^{\prime})$ there will be no Sagnac effect. In general, for non-zero $q$ (non-mid point measurement) there will be a delay in arrival times of two oppositely moving light rays. Such a delay, however, does not depend on the velocity of the observer. The same conclusion can be reached in the Lab (un-primed) frame also.

Next, we consider the detector is moving, but the mirrors are at rest in the Lab frame. This situation is essentially similar to the rectangular closed path case considered in \cite{tartaglia2015sagnac}. Here, the trajectory for the light rays that starts moving initially towards the negative x-axis is $OBOAD_{1}$ , where $D_{1}$ is the position of the detector when the light ray reaches the detector. The total time taken by the stated light rays to reach the detector is
\begin{equation}
\Delta t_{-}= \frac{4L-q}{c+v}. 
\end{equation}

For reaching the detector at D, the path of the light rays that propagate towards positive x-axis initially is $OAOBOD_{2}$ and the total time taken in the process is 
\begin{equation}
\Delta t_{+}= \frac{4L-q}{c-v}. 
\end{equation}

The time difference of arrival of two opposite directed light rays at the detector is 
\begin{eqnarray}
\Delta t = \Delta t_+ - \Delta t_- = \frac{8Lv+2qc}{c^2-v^2}. 
\end{eqnarray}

Note that the measurement will be made by the moving observer. Hence, applying the Lorentz transformations the time delay to be measured by the moving observer, as inferred from the Lab frame will be 

\begin{equation}
\Delta t_{obs} = \frac{8Lv+2qc}{\gamma (c^2-v^2)}. 
\end{equation}

When $q=0$, one gets the usual generalized linear Sagnac (like) effect. On the other hand, if $q=-4Lv/c$, there will be no Sagnac delay, implying that the measurements of the arrival times are made exactly at point O which is the mid-point of the light trajectories (between the mirrors). 

In the primed frame, the detector is at rest. The mirrors are moving with a velocity $v$ towards the negative x-direction. The light path for the rays that starts moving initially towards the negative x-axis is $O^{\prime}B_1^{\prime}OA_2^{\prime}O^{\prime}$, where $B_1^{\prime}$ is the position of the mirror when light reached the reflector, which was initially at $B$. At that moment, the mirror $A$ was at $A_1^{\prime}$ position. After reflecting by the mirror at $B_1^{\prime}$ the light ray will move towards positive x-axis direction and reach the reflector $A$ at $A_2^{\prime}$ position. Since $O^{\prime}B_1^{\prime}$ not equals to $O^{\prime}A_2^{\prime}$, $O^{\prime}$ is not the mid-point of the light path. The detector bound observer will find that the lengths contracted according to STR. Accordingly, in the primed frame the  difference in time of arrival of two opposite directed light rays at the detector will be 

\begin{equation}
\Delta t^{\prime}_{obs} = \frac{8Lv+2qc}{\gamma (c^2-v^2)}. 
\end{equation}

which is exactly what the Lab frame is interpreted. The difference in arrival times of the two oppositely moving light beams in the lab frame and the frame attached to the detector are, thus, simply connected by special relativistic time dilation as expected. 

\section{On the original Sagnac experiment}
\begin{figure}
	\begin{center}
		\begin{tikzpicture}[scale=0.75]
		\draw[thick] (1,0) arc (0:360:4cm);
		\draw[->] (0,0) arc (0:330:3cm);
		\draw[blue,<<-] (-.30,0) arc (0:400:2.7cm);
		\draw[red,->] (-1,0) arc (0:-120:2.0cm);
		\begin{scope}[xshift=1cm, yshift =-1.6cm]
		\draw[dashed](0,0)--(-2,2);
		\end{scope}
		\begin{scope}[xshift=1cm, yshift =-2.9cm]
		\draw[dashed,red](0,0)--(-2,2);
		\end{scope}
		\begin{scope}[xshift=.5cm, yshift =-3.5cm]
		\draw[dashed,green](0,0)--(-2,2);
		\end{scope}
		\node[right] at (0,0) {$O$};
		\node[right] at (0,-.6) {$P$};
		\node[right,red] at (0,-1.7) {$D_1$};
		\node[right,green] at (0,-3) {$D_2$};
		\node[right,red] at (-3.0,-1.3) {$\omega$};		
		\end{tikzpicture}
	\caption{Original Sagnac Modified}
	\label{fig:original}
	\end{center}
\end{figure}
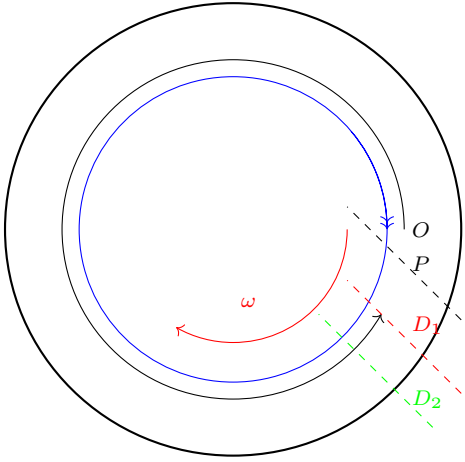

Now finally let us consider the original Sagnac experiment as displayed in fig. \ref{fig:original}. Here two light rays start to move, say at time $t_o$, from the point $O$ in opposite directions, clockwise and anti-clockwise along the perimeter of a circle of radius $r$. At that instant, an observer with an interferometer is placed at $P$, where the arc $OP = p$. The whole set up is placed on a rotating disk/platform having angular velocity $\omega$ (in lab frame) with the center of the disk coincides with that of the circular path of the light rays. 

In the Lab frame, the difference in arrival time of clockwise and anticlockwise light rays in the interferometer will be 

\begin{equation}
\Delta t = \frac{4\pi r v + 2pc}{c^2-v^2}. 
\label{Eq:rl}
\end{equation}

where $v=r \Omega$, $\Omega$ being the angular velocity of the rotating platform. Translating the above time delay to the observation by the rotating observer is not straightforward. Some scientists suggest that relativistic kinematic transformations are valid not only between inertial frames with uniform relative velocity but also for reference frames undergoing acceleration \cite{Malykin2000TheSE}. Following (i.e. essentially assuming that the Lorentz transformations hold between the Lab frame and the rotating observer) \cite{Malykin2000TheSE}, the difference in arrival time of clockwise and anticlockwise light rays in the interferometer of the rotating observer will be
      
\begin{equation}
\Delta t^{\prime} = \frac{4\pi r v + 2pc }{\gamma (c^2-v^2)}. 
\label{Eq:rot1}
\end{equation}

The above expression, which is obtained from Eq. \ref{Eq:rl} taking in to account the time dilation effect, generalizes the one given in \cite{Malykin2000TheSE} for arbitrary initial position of the observer. Again for p=0, one gets the usual expression for the Sagnac delay. For $p=-2\pi r v/c$, i.e. when the measurements are made at the point of initiation of the journey there will be no Sagnac delay.   

Several researchers prefer to explain the Sagnac effect in the purview of general theory of relativity (GR). Langevin first explained the Sagnac effect from GR consideration. The basic approach in GR based explanation is to compute the difference of propagation times in reaching the detector between counter-propagating waves in the space-time metric that effectively represents the frame of reference attending rotation. In the laboratory frame the space-time interval in cylindrical coordinates $(t, r, \phi, z)$ is given by

\begin{equation}
ds^2= c^2 dt^2 - dr^2 - r^2 d\phi^2 - dz^2
\end{equation}

Now when the platform starts rotating with uniform rotating velocity $\Omega$ along z-axis, the space time interval associated with the platform $(t^{\prime} = t, r^{\prime}=r, \phi^{\prime} = \phi + \Omega t, z^{\prime} = z)$ \cite{landau1975classical}

\begin{equation}
ds^2= (c^2 - r^2 \Omega^2) dt^2 - 2 \Omega r^2 d\phi dt - dr^2 - r^2 d\phi^2 - dz^2
\label{Eq:rot}
\end{equation}

The space time curvature vanishes for the above space time metric implying that there is no gravitational field associated with the metric. For the circular ($dr=0$) light trajectory $(ds=0)$ at $r\phi$ plane the eq. (\ref{Eq:rot}) gives

\begin{equation}
dt^{\prime}_{cl,acl}= \frac{\pm r d\phi}{c \mp r\Omega}
\label{Eq:root}
\end{equation}

The subscript $cl$ and $acl$ denote clockwise and anticlockwise respectively. The total propagation time for clockwise light rays to reach the observer

\begin{equation}
\Delta t^{\prime}_{cl} = \int_{-p}^{2\pi} dt^{\prime}_{cl} =  \frac{2\pi r +p}{c-r\Omega}
\label{Eq:rotcl}
\end{equation}
 
and that for anticlockwise light rays
\begin{equation}
\Delta t^{\prime}_{acl} = \int_{-p}^{-2\pi} dt^{\prime}_{cl} =  \frac{2\pi r -p}{c+r\Omega}
\label{Eq:rotcl1}
\end{equation}

Hence the difference in arrival times between the counter-propagating light rays is exactly the same to eq. (\ref{Eq:rot1}) as obtained in the Lab frame. However, one may notice that the above derivation does not give the time dilation effect. In his review article Post \cite{post1967sagnac} argued that the time coordinate should transform as $t^{\prime}= \gamma t$ while switching over from Lab frame to stationary frame which leads to the time dilation effect.    

A worthwhile point to be noted that the metric given in eq. (\ref{Eq:rot}) is derived from the Lab frame space time metric; it is the metric of the rotating frame according to a Lab frame observer. Though mathematically it is fine but the physical understanding of the effect from the standpoint of an observer attending the rotation remains difficult. When $p=0$, i.e. the observer at the point of splitting remains on that point in the rotating frame, why clockwise and anti-clockwise light will take different times to reach the observer?         

To understand the point mentioned above let us first consider a Sagnac like thought experiment involving rectangular trajectories of light rays in the Lab (inertial) frame as discussed in \cite{tartaglia2015sagnac}. Say, $ABCD$ is a square/rectangle in an inertial frame $S$. Using a beam splitter, a light signal is directed to move along the paths $ABCDA$ and $ADCBA$ (reflectors placed at points $A, B, C, D$).  Both the light rays initially start their motion from the same point A at the same time. They are expected to meet at $A$, where an interferometer is placed, at the same instant. We allow the interferometer, which was initially at point $A$ at the time of the two light beams started their journey along $ABCDA$ and $ADCBA$, to move along $AB$ with a constant velocity $v$. For simplicity of the calculation the length of the sides and the velocity $v$ are so chosen that both the clockwise and anticlockwise light beams reach the interferometer before it (interferometer) crosses the $B$ point i.e. $AB/v > (3AB+2BC)/c$ which implies $v < c/(3+2\xi)$ where $\xi = BC/AB$. 

\subsection{From the Lab frame}
\begin{center}
\begin{figure}
\centering
\begin{tikzpicture}[scale=0.66]
\draw[thick, ->] (0,0)--(0,4) node[anchor =north west]{D};
\draw[thick, ->] (0,4)--(4,4) node[anchor =north east]{C};
\draw[thick, ->] (4,4)--(4,0) node[anchor =south east]{B};
\draw[thick, ->] (4,0)--(1.6,0) node[anchor =south]{I};
\draw (4,0)--(0,0) node[anchor =south west]{A};
\fill[blue!40!white] (1.5,0) circle (1mm);
\end{tikzpicture}
\qquad
\begin{tikzpicture}[scale=0.66]
\draw[thick, ->] (0,0)--(4,0) node[anchor =south east]{B};
\draw[thick, ->] (4,0)--(4,4) node[anchor =north east]{C};
\draw[thick, ->] (4,4)--(0,4) node[anchor =north west]{D};
\draw[thick, ->] (0,4)--(0,0) node[anchor =south west]{A};
\draw[thick, ->] (0,0)--(1.4,0) node[anchor =south]{I};
\fill[blue!40!white] (1.5,0) circle (1mm);
\end{tikzpicture}

\caption{As seen from Lab frame: (a) clockwise (b) counter-colckwise}
\label{fig:lab}
\end{figure}
\end{center}

We shall now explore what will be the observations of the lab frame and the observer attached with the interferometer. From the viewpoint of the lab frame, the light rays that travel along ADCBI path will reach the interferometer (I) earlier than that propagate along ABCDAI (please refer to figure 1(a)). Denoting the length of the sides AB=CD=p and BC=DA=q, 
the difference in arrival times of the anticlockwise and clockwise moving light beams is
\begin{equation}
\Delta \xi = \frac{4(p+q)v}{c^2-v^2}
\end{equation}
which is similar to the Sagnac effect; the only difference is that in the Sagnac effect the perimeter of a circle occurs instead of a rectangle, as the light paths in the Sagnac experiment are circular~\cite{tartaglia2015sagnac}. 

\begin{figure}
\centering
\begin{tikzpicture}[scale=0.66]
\draw (0,0) node[anchor =south west ]{A,I} -- (4,0)node[anchor =south east]{B} -- (4,4) node[anchor =north east]{C} -- (0,4)node[anchor =north west]{D} -- (0,0);
\draw[thick, red, ->] (0,0) -- (-.5,4) node[anchor =north east]{D1};
\draw[thick, red, ->] (-.5,4)--(3.2,4) node[anchor =north east]{C1};
\draw[thick, red,->] (3.2,4)--(2.9,0) node[anchor =south east]{B1};
\draw[thick, red,->] (2.9,0)--(0.1,0);
\fill[blue!40!white] (0,0) circle (1mm);
\end{tikzpicture}
\qquad
\begin{tikzpicture}[scale=0.66]
\draw (0,0) node[anchor =south west ]{A,I} -- (4,0)node[anchor =south east]{B} -- (4,4) node[anchor =north east]{C} -- (0,4)node[anchor =north west]{D} -- (0,0);
\draw[thick, red, ->] (0,0)--(3.5,0) node[anchor =south east]{B1};
\draw[thick, red, ->] (3.5,0)--(3,4) node[anchor =north east]{C1};
\draw[thick, red, ->] (3,4)--(-.5,4) node[anchor =north east]{D1};
\draw[thick, red, ->] (-.5,4)--(-1.5,0) node[anchor =south west]{A1};
\draw[thick, red, ->] (-1.5,0)--(-.10,0);
\fill[blue!40!white] (0,0) circle (1mm);
\end{tikzpicture}
\caption{As seen from observer frame: (a) clockwise (b) counter-clockwise}
\label{fig:obs}
\end{figure}
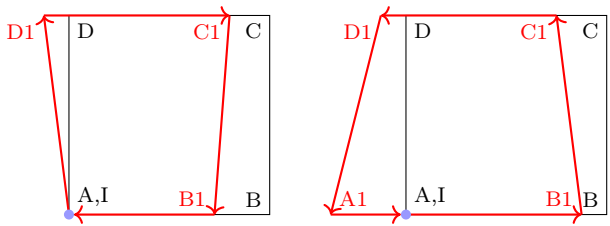

Now we consider the viewpoint of the observer $O’$ attached with the interferometer. For $O’$ the points $A, B, C$ and $D$ are moving with a constant velocity $–v$. In this frame the length $AB$ and $DC$ will be contracted following the Lorentz contraction formula whereas the length $BC$ and $AD$ will remain unaltered. The light trajectories for clockwise and anticlockwise motion as will perceive by $O’$ are shown in fig.(\ref{fig:obs}).
  
Therefore, the arrival time difference of the anticlockwise and clockwise moving light beams as be viewed by $O’$ is:
\begin{equation}
\Delta \xi ' = \frac{4(p+q)v}{\gamma (c^2-v^2)}
\label{eq:6}
\end{equation} 

Here also the difference in arrival times of the anticlockwise and clockwise moving light beams in the lab frame and the frame attached to the detector are simply connected by special relativistic time dilation as expected. The most striking point of the above Sagnac kind thought experiment involving rectangular trajectories (in Lab frame) is that in the frame attached with the interferometer, the geometries of the travelling paths for the clockwise and the anticlockwise light rays are not rectangular; they trace two different unequal paths while moving from the point $A$ to the Interferometer as shown in fig. (\ref{fig:obs}).

So a relevant question is whether in original Sagnac experiment the light trajectories are circular in the frame attending the rotation? In the light of the so called Ehrenfest paradox, a circle in the Lab frame is unlikely to be perceived as a circle by the observer attending the rotation.   

\section{Discussion}
We conclude that the origin of the Sagnac Delay or the phase difference in Sagnac or Sagnac-like experiments is the non-mid-point measurement of arrival times of counter-propagating waves leading to unequal path lengths traversed by the oppositely directed light rays in reaching the interferometer. It does not depend exclusively on the rotation, as correctly pointed out in \cite{tartaglia2015sagnac}. The synchronization issues in the rotating frame cannot be the cause as we have seen that zero Sagnac delay is possible in the Sagnac experiment depending on the arrival times measurement location. The relative motion between the detector and the reflectors is also not the reason for the Sagnac delay.  

In the lab frame, the non-mid point position of the observer and thereby inequality of the path lengths of the two oppositely directed light rays in the Sagnac experiment is straightforward. However, from the standpoint of the rotating observer, the issue is quite difficult to understand.

So we have considered a linear version of the Sagnac experiment, where, because of the linear constant relative speed of the interferometer and mirrors, one can easily deduce the light paths in the interferometer frame. In the frame attached with the interferometer, the oppositely directed light rays trace unequal paths while moving from the point $O$ to the interferometer, if the observation is conducted any other point than $O$. So from both the concerned frames, the underlying reason for the observed phase difference is the same, the non-mid point observation. If the detector has zero velocity but the measurement is made at any non-mid-point, there will be a Sagnac kind time delay but obviously, it (delay) will not depend on the velocity. Such a linear version is not exactly the same as the Sagnac experiment. In the proposed linear case there is a relative velocity between the detector and the reflectors and the distance between the detector and the reflectors continuously alters, unlike the Sagnac experiment where such distances always remain the same, at least in the lab frame. The gedanken original Sagnac kind experiment involving rotation also validates the non-mid-point measurement as the root cause of the Sagnac effect in the Lab frame.      

An interesting question is whether a time dilation effect will be present in a Sagnac effect for the rotating observer. The Lorentz $\gamma$-factor occurs for linear constant relative speed between two inertial frames. For rotating motion, there may be some modification in that factor \cite{selleri96,grcs}. This has been discussed in detail in view of the Ehrenfest paradox using LSE in Ref.~\cite{grcs, biplabthesis}. The proposed thought experiment cannot say anything about the appearance of the Lorentz $\gamma$-factor in a rotating frame. An experimental determination of this factor would be of considerable interest.


\end{document}